\documentclass[mathleft
]{an}
\usepackage{graphicx}
\usepackage{times}
\usepackage{subfigure}
\def\apj{ApJ}
\def\mnras{MNRAS}
\def\apjl{ApJ}

\def\aap{A\&A}

\def\aj{AJ}

\def\physrep{Phys. Rep.}
\def\apjs{ApJS}
\def\nat{Nature}

\def\araa{Ann. Rev. A\&A}
\def\aapr{A\&ARv}

\def\Msun{\hbox{$\thinspace M_{\odot}$}}

\def\xmm{{\it XMM-Newton}}
\newcommand{\xmmn}{{\it XMM-Newton~\/}}

\def\Mbh{\hbox{$M_{\rm{BH}}$}}

\def\Mdot{\hbox{$\dot M$}}

\def\mdotedd{\hbox{$\dot m_{\rm Edd}$}}

\def\Ledd{\hbox{$L_{\rm Edd}$}}
\def\xmm{{\it XMM-Newton}}
\newcommand{\ms}{MS 2254.9--3712}
\newcommand{\rej}{RE J1034+396}

\def\gsim{\mathrel{\hbox{\rlap{\hbox{\lower4pt\hbox{$\sim$}}}\hbox{$>$}}}}
\def\lsim{\mathrel{\hbox{\rlap{\hbox{\lower4pt\hbox{$\sim$}}}\hbox{$<$}}}}

\begin{document}

\Pagespan{000}{}
\Yearpublication{0000}%
\Yearsubmission{0000}%
\Month{00}%
\Volume{000}%
\Issue{00}%
\DOI{This.is/not.aDOI}%


\title{Quasi periodic oscillations in active galactic nuclei}

\author{William Alston\inst{1}\fnmsep\thanks{Corresponding author:
  \email{wna@ast.cam.ac.uk}\newline}
\and  ~Andy Fabian~\inst{1}
\and  ~Julija Markevi{\v c}i{\= u}t{\.e}\inst{2}
\and  ~Michael Parker\inst{1}
\and  \\Matt Middleton\inst{1}
\and  Erin Kara\inst{1}
}
\titlerunning{Instructions for authors}
\authorrunning{Alston et al }
\institute{
Institute of Astronomy, Madingley Rd, Cambridge, CB3 0HA, UK
\and 
Department of Applied Mathematics and Theoretical Physics, Centre for Mathematical Sciences, Wilberforce Rd, Cambridge CB3 0WA
}
\received{3 September 2015}


\abstract{%
Quasi-periodic oscillations (QPOs) are coherent peaks of variability power observed in the X-ray power spectra (PSDs) of stellar mass X-ray binaries (XRBs).  A scale invariance of the accretion process implies they should be present in the active galactic nuclei.  The first robust detection was a $\sim 1$\,hr periodicity in the Seyfert galaxy \rej~from a $\sim 90$\,ks \xmmn observation, however, subsequent observations failed to detect the QPO in the $0.3-10.0$\,keV band.  In this talk we present the recent detection of the $\sim 1$\,hr periodicity in the $1.0-4.0$\,keV band of 4 further low-flux/spectrally-harder observations of \rej~(see Alston et al 2014).  We also present recent work on the discovery of a QPO in the Seyfert galaxy, \ms, which again is only detected in energy bands associated with the primary power-law continuum emission (Alston et al 2015).  We conclude these features are most likely analogous to the high-frequency QPOs observed in XRBs.  In both sources, we also see evidence for X-ray reverberation at the QPO frequency, where soft X-ray bands and Iron K$\alpha$ emission lag the primary X-ray continuum.  These time delays may provide another diagnostic for understanding the underlying QPO mechanism observed in accreting black holes.}

\maketitle

\section{Introduction}

One of the most important breakthroughs in the study of stellar mass black holes was the discovery of high frequency quasi-periodic oscillations (HFQPOs; with $\nu_{\rm qpo} \gsim 40$\,Hz) in Galactic black hole X-ray binaries (XRBs), with $\Mbh \sim 10 \Msun$ (Morgan, Remillard \& Greiner 1997; Remmillard et al 1999; Strohmayer et al 2001; Remillard et al 2002, 2003; van der klis 2006; Remillard \& McClintock 2006).  Sixteen years of RXTE observations have yielded detections in a handful of sources (see Belloni, Sanna \& Mendez 2012; Altamirano \& Belloni 2012; Belloni \& Altamirano 2013a,b, and references therein).

These are the fastest coherent variability signatures observed from XRBs, with frequencies close to Keplerian frequency of the innermost stable circular orbit (ISCO).  HFQPOs are therefore expected to carry information on the strongly curved spacetime close to the black hole, providing constraints on the two fundamental properties of black holes: mass and spin. 

Despite their importance, the exact mechanism responsible for producing HFQPOs remains one of the biggest challenges in contemporary astrophysics, with varying scenarios for the origin of the oscillation presented in the literature (e.g. Milsom \& Taam 1997; Nowak et al 1997; Wagoner 1999; Stella et al 1999; Abramowicz \& Kluzniak 2001; Rezzolla et al 2003; Das \& Czerny 2011, see also Motta, S., this volume).

If strong gravity dominates the accretion process, a scale invariance implies that HFQPOs should also be present in Active Galactic Nuclei (AGN; $\Mbh \gsim 10^{6} \Msun$).  With a black hole (BH) mass ratio $M_{\rm BHB} / M_{\rm AGN} = 10^{-5}$ we expect $\nu_{\rm QPO} \gsim 5 \times 10^{-3}$\,Hz (i.e. timescales of $\gsim 200$\,s) in AGN, well within the temporal passband of e.g. \xmm.  AGN have higher counts per characteristic timescale, allowing us to potentially study individual QPO periods, providing a better opportunity to understand this phenomenon.

QPOs are notoriously difficult to detect in AGN, primarily due to insufficient observation lengths currently available with \xmmn (Vaughan \& Uttley 2005, 2006).  Many reported detections have been disfavoured due to inaccuracy of modelling the underlying red noise continuum (Vaughan 2005; Vaughan \& Uttley 2005, 2006; Gonzalez-Martin \& Vaughan 2012).  A $\sim 1$\,hr periodicity in the Seyfert galaxy \rej~was the first robust detection of a QPO in an AGN (Geirlinski et al 2008, see also Vaughan 2010).

Accreting BHs also display {\it hard} lags at low frequencies --- where variations in harder energy bands are delayed with respect to softer energy bands (e.g. Alston et al 2014 and references therin).  The leading model for the origin of the hard lags is the radial propagation of random accretion rate fluctuations through a stratified corona (e.g. Arevalo \& Uttley 2006 and references therein).

Soft X-ray lags at higher frequencies has now been observed in $\gsim 20$ AGN (e.g. Fabian et al 2009; Emmanoulopoulos et al 2011; Zoghbi 2011; Alston et al 2013; Cackett et al 2013; De Marco et al 2013; Kara et al 2013; Alston et al 2014).  A picture is emerging where the \emph{reverberation} signal is produced when the primary X-ray emission is reprocessed by the inner accretion disc (see Uttley et al 2014 for a review).  High frequency iron K$\alpha$ lags have also been observed, adding more weight to the inner reverberation scenario (e.g. Zoghbi et al 2012; Zoghbi et al 2013; Kara et al 2013).  See talks by Cackett, E., De Marco, B., Ingram, A., Kara, E., Uttley, P. in this volume for more on time delays in accreting sources.

The time lags of HFQPOs in XRBs have also been studied, with Mendez et al (2013) recently carrying out a systematic study in 4 sources.  They found both hard and soft lags are observed at the QPO and harmonic frequencies.  The interpretation of these HFQPO time lags is still unclear, however, they provide an extra diagnostic for understanding their physical origin.

In this talk, I will present recent work on HFQPOs in two AGN: \rej~and \ms~, as well as their associated variability properties.

\section{Recent QPO detections in AGN}
\subsection{RE J1034+396}

RE J1034+396 is a nearby ($z = 0.042$) NLS1 galaxy with $\Mbh \sim 2 \times 10^{6} \Msun$ and thought to be accreting at or above the Eddington rate ($L/\Ledd \sim 1$; e.g. Jin et al 2012).  This led Middleton \& Done (2010) and Middleton et al (2011, hereafter M10 and M11, respectively) to conclude that the QPO in \rej~is an analogue of the 67\,Hz QPO in the super-Eddington BHB GRS 1915+105.  Four further \xmmn observations of \rej~were analysed in M11 in an attempt to confirm the QPO in this source.  Restricting the PSD analysis to the $0.3-10.0$\,keV band revealed no sign of the QPO at any frequency, leading M11 to conclude the QPO was a transient feature.  However, similarities between the covariance spectra (Wilkinson \& Uttley 2009) were seen between the original $90$\,ks observation and two other observations.  Further attempts at understanding the QPO in \rej~have so far been limited to the one $90$\,ks observation (Czerny et al 2010; M10; Czerny et al 2012; Hu et al 2014).

\subsubsection{Power-spectrum analysis}

In this work we present an energy resolved PSD analysis of \rej, see Alston et al 2014b (hereafter, A14) for full details.  We used all eight archival \xmmn observations, spanning 2002 to 2011.  Light curves were extracted following standard methods in a soft (0.3-1.0\,keV) and hard (1.0-4.0\,keV) bands (see Figure 1 in A14).  The hard band was chosen to provide a high S/N light curve of the primary continuum.  PSDs were estimated by calculating the periodogram (e.g. Priestley 1981; Percival \& Walden 1993; Vaughan et al 2003a).  We fitted the PSDs using the maximum likelihood method described in Vaughan 2010, see A14 for fit details.

\begin{figure*}
\centering
\mbox{\subfigure{\includegraphics[width=0.3\textwidth,angle=90]{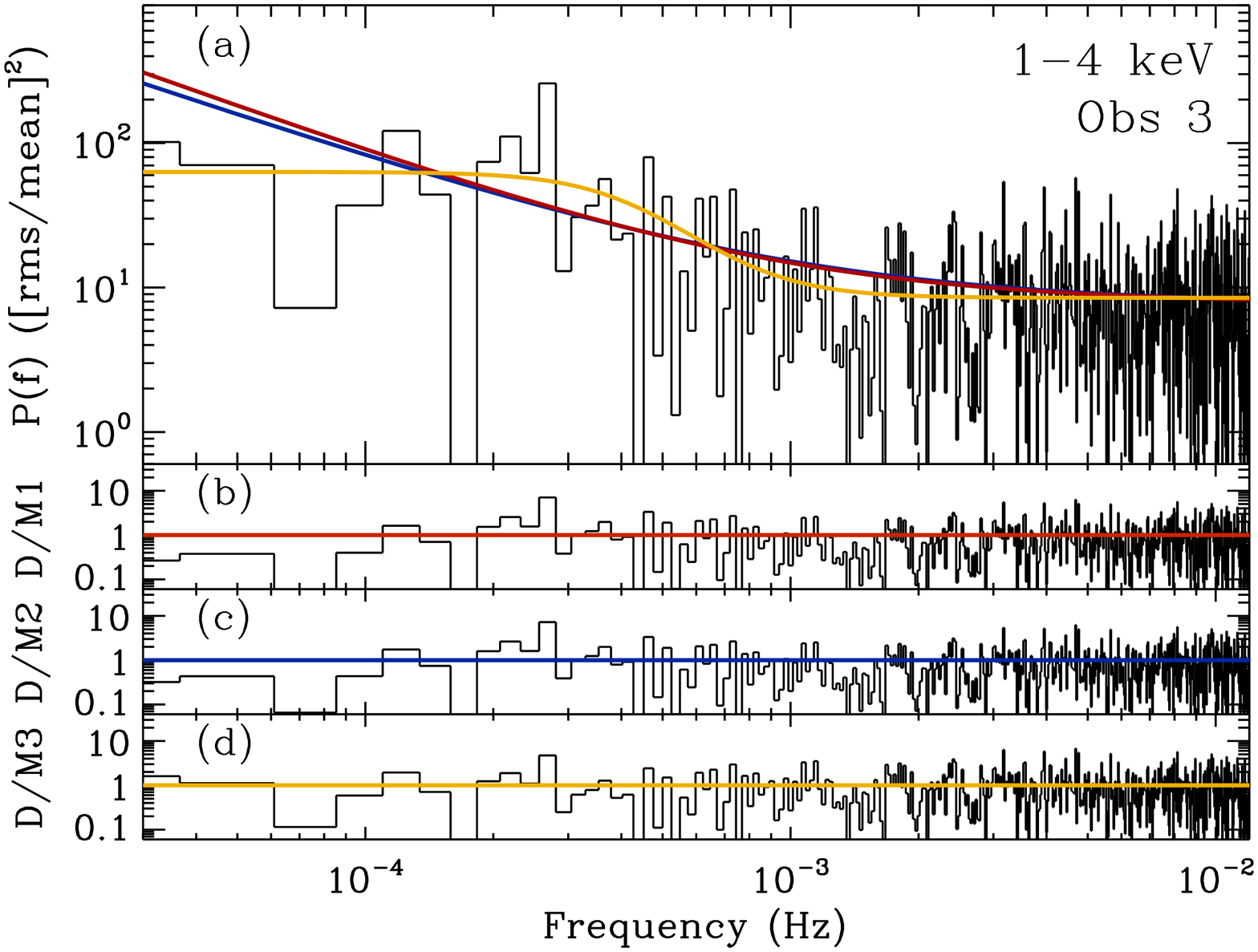}}
\hspace{15pt}
\subfigure{\includegraphics[width=0.3\textwidth,angle=90]{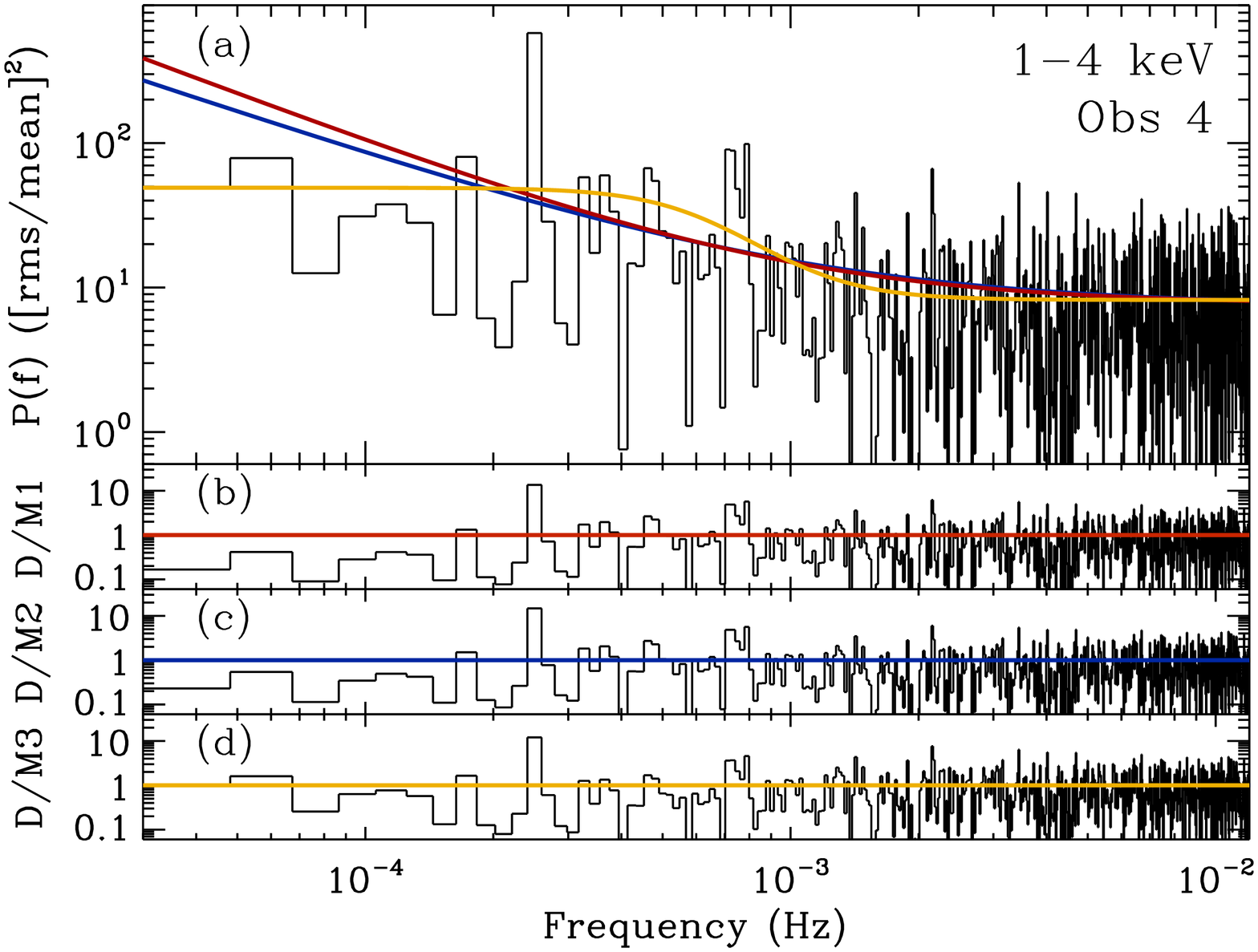}}
}\\
\mbox{\subfigure{\includegraphics[width=0.3\textwidth,angle=90]{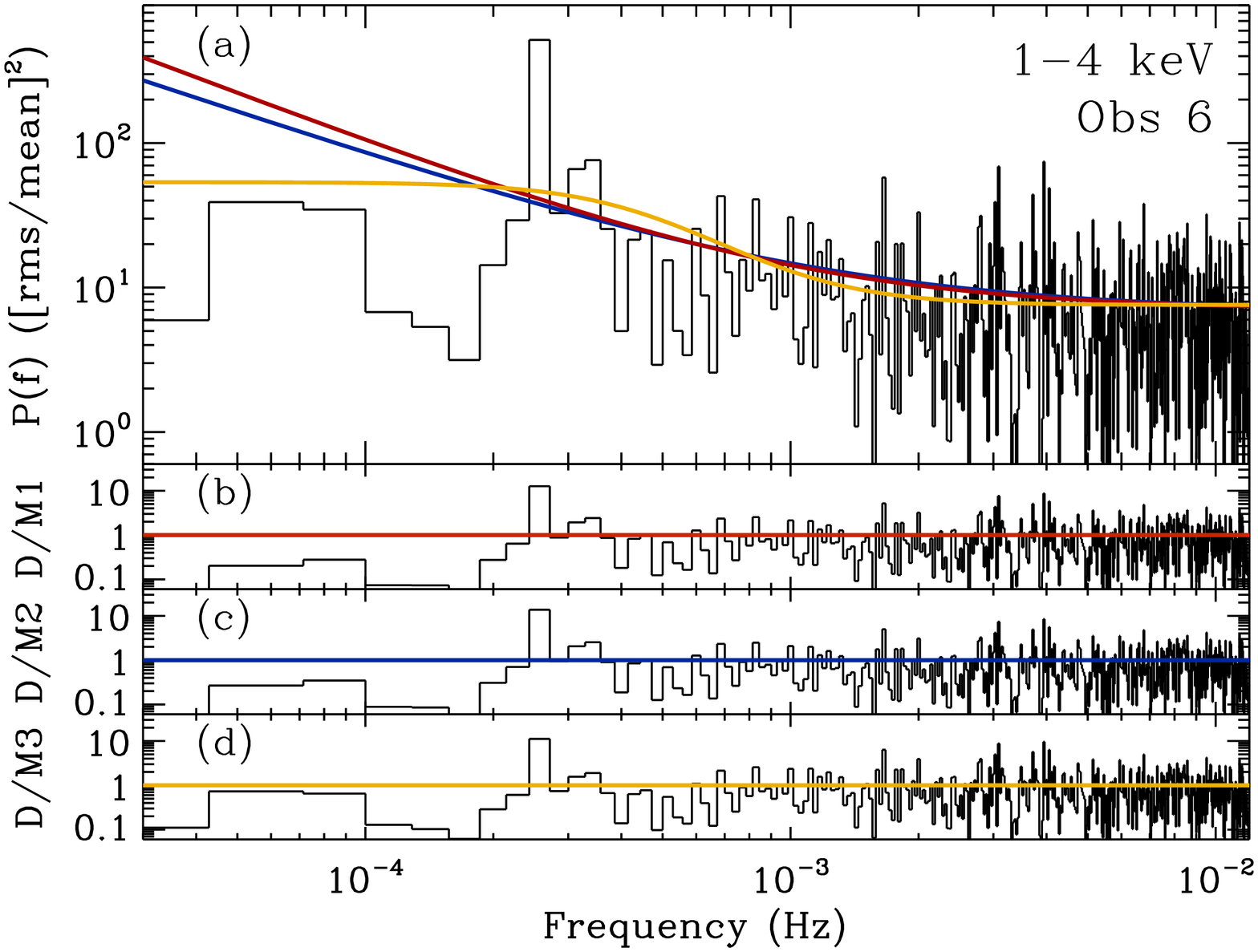}}
\hspace{15pt}
\subfigure{\includegraphics[width=0.3\textwidth,angle=90]{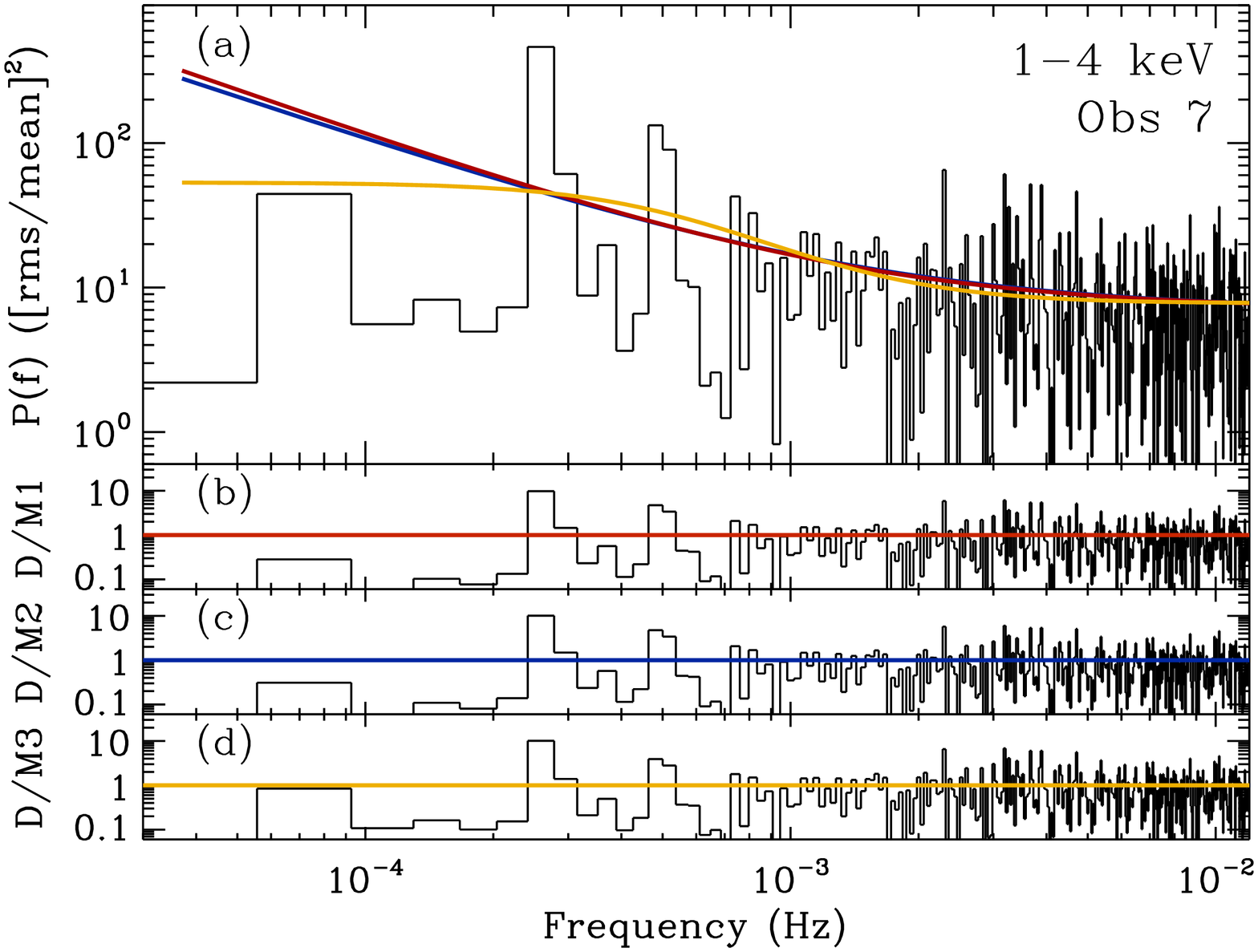}}
}\\
\caption{The model fits to the $1.0-4.0$\,keV PSDs of Obs3,4,6 and 7 in \rej~(see A14 for observation details).  In panel (a) is the data and PSD model fits for Model 1 (red), Model 2 (blue) and Model 3 (yellow).  Panels (b), (c) and (d) show the data/model residuals for models 1,2 and 3 respectively.  The frequency of the QPO has remained the same over 5 years of observations.  A tentative harmonic component is seen in Obs 7.  Figures reproduced from A14.}
\label{fig:rejpsd}
\end{figure*}

We used two simple continuum models that are commonly used to fit AGN PSDs (e.g. Uttley et al 2002; Vaughan \& Fabian 2003; Vaughan et al 2003a,b; McHardy et al 2004; Gonzalez-Martin Vaughan 2012).  Model 1 is the simplest model consisting of a power law plus constant:

\begin{equation}
\label{eqn:pl}
   P(\nu) = N \nu^{- \alpha} + C
\end{equation}

\noindent where $N$ is the normalisation term and $C$ is a non-negative constant used to model the Poisson noise level.  Model 2 is a bending power-law (e.g. McHardy et al 2004):

\begin{equation}
\label{eqn:bendpl}
   P(\nu) = \frac{N \nu^{{\alpha}_{\rm low}}}{1 + (\nu / \nu_{\rm bend})^{{\alpha}_{\rm low}-{{\alpha}_{\rm high}}}} + C
\end{equation}

\noindent where $\nu_{\rm bend}$ is the bend frequency, ${\alpha}_{\rm low}$ and ${\alpha}_{\rm high}$ are power-law slopes below and above $\nu_{\rm bend}$ respectively.  In Model 2 we fixed ${\alpha}_{\rm low} = 1$, which is the typical value found from long-term X-ray monitoring studies (e.g. Uttley et al 2002; Markowitz et al 2003; McHardy et al 2004; Gonzalez-Martin Vaughan 2012).  A variant of this model (Model 3) with ${\alpha}_{\rm low} = 0$ is used as the significance of QPO features has a strong dependence on the continuum modelling (e.g. Vaughan \& Uttley 2005, 2006).  The model fits to the 4 low-flux/spectrally harder observations are shown in Fig.~\ref{fig:rejpsd}.

Using a likelihood ratio test we find that the simple model 1 is prefered all the time for all energy bands. A significant outlier (i.e low $p$-value) is found in the hard band (1.0-4.0\,keV) in the five observations which have a lower flux and are spectrally harder, including the original $90$\,ks observation.  No significant outliers are observed in the soft or total bands in these 5 observations, with the exception of the $90$\,ks observation.  No significant outliers are observed at any energy in the two higher flux / spectrally softer observations.  The fractional variability amplitude of the feature is $\sim 5 \%$ in each of the 5 observations.  A quality factor $Q = \nu / \Delta \nu \gsim 10$ is observed in the 5 detections.

\subsection{\ms}

\ms~is a nearby ($z = 0.039$; Stocke 1991) `unabsorbed' ($N_{\rm H} < 2 \times 10^{22}~{\rm cm}^{-1}$; Grupe et al 2004) narrow line Seyfert 1 (NLS1) galaxy, with X-ray luminosity log~($L_{\rm X}) = 43.29~{\rm erg~s^{-1}}$.  The central BH mass in \ms~is estimated as $\Mbh \sim 4 \times 10^{6}$ using the empirical $R_{\rm BLR}-\lambda L_{\lambda}(5100 {\rm \AA})$ relation (Grupe et al 2004; Grupe et al 2010) and is estimated as $\Mbh \sim 10^{7}$ using the $\Mbh - \sigma$ relation (Shields et al 2003).  The Eddington rate is estimated as $L_{\rm Bol}/\Ledd = 0.24$ using $\lambda L_{\lambda}(5100 {\rm \AA})$ (Grupe et al 2004; Grupe et al 2010).  However, Wang et al 2003 suggest \ms~is accreting at super-Eddington rate ($\Mdot / \mdotedd > 1$).

\begin{figure}
\centering
\includegraphics[width=0.3\textwidth,angle=90]{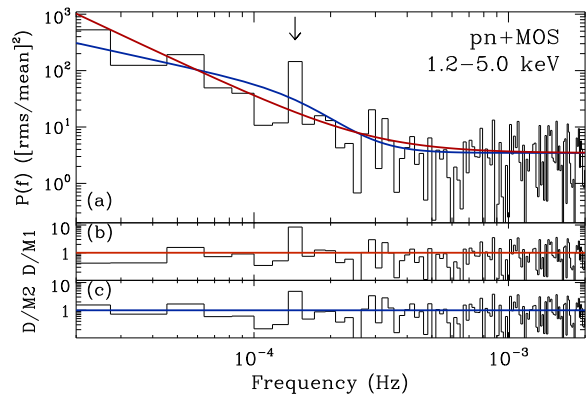}
\caption{The 1.2--5.0\,keV band PSD and model fits are shown in panel (a), for model 1 (red) and  model 2 (blue).  The data/model residuals for models 1 and 2 are shown in panels (b) and (c), respectively. Figure reproduced from A15.}
\label{fig:mspsd}
\end{figure}

We model the PSD in \ms~in a similar manner to \rej~(see Alston et al 2105, hereafter A15).  Fig.~\ref{fig:mspsd} shows the model fits to the $1.2 - 5.0$\,keV band, where a significant outlier can be seen at $\sim 1.5 \times 10^{-4}$\,Hz ($\sim 2$\,hrs).  No significant features are observed in other energy bands.  The fractional variability amplitude of the feature is $\sim 6 \%$ and the quality factor $Q = 8$, consistent to what is observed in \rej.

\section{Frequency resolved time delays}

In this section we explore the frequency-dependent time delays between different energy bands in both \rej~and \ms.  This allows us to study any reprocessed emission that is responding to the QPO modulation.

\begin{figure}
\centering
\subfigure{\includegraphics[width=0.4\textwidth,angle=0]{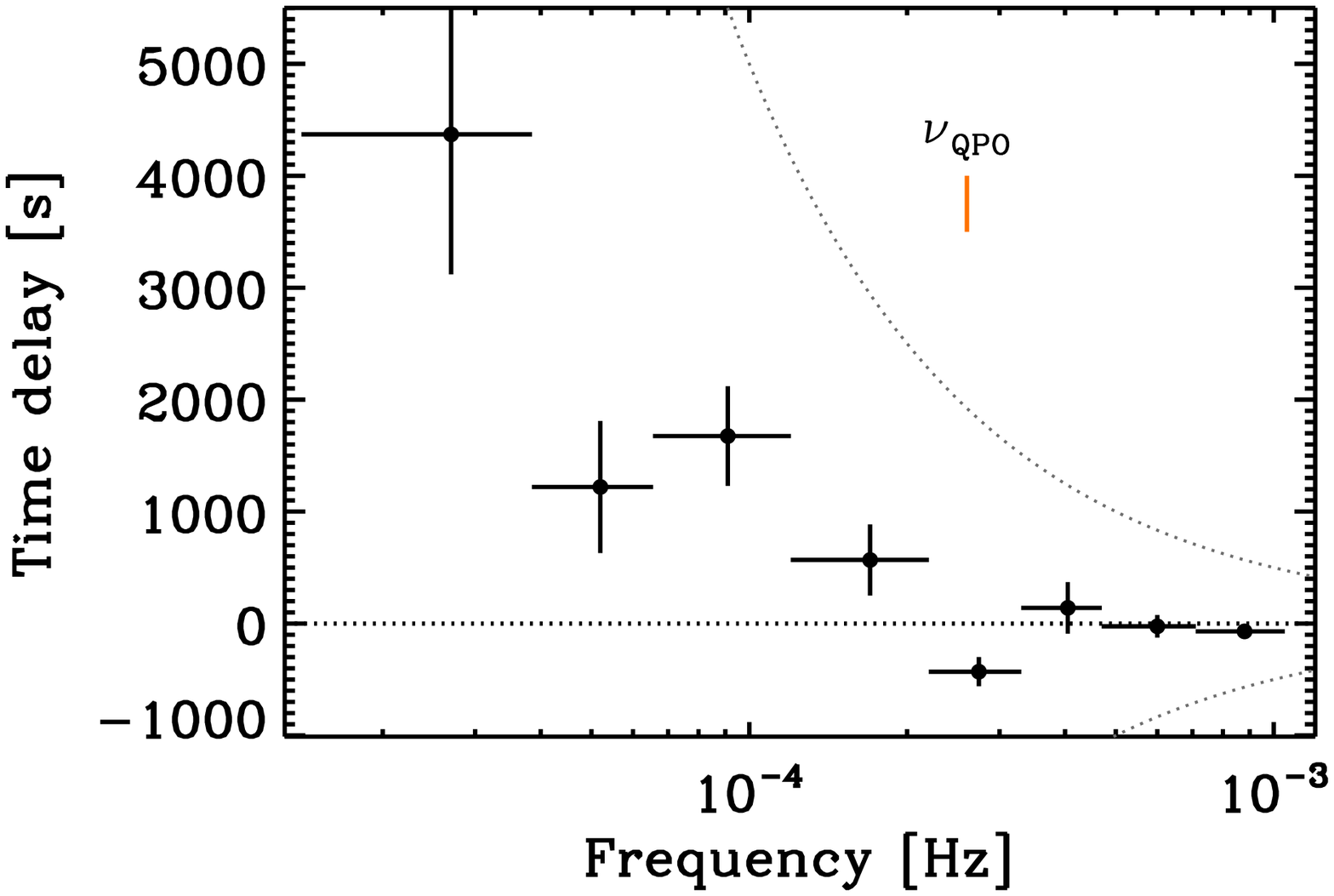}}
\\
\subfigure{\includegraphics[width=0.42\textwidth,angle=0]{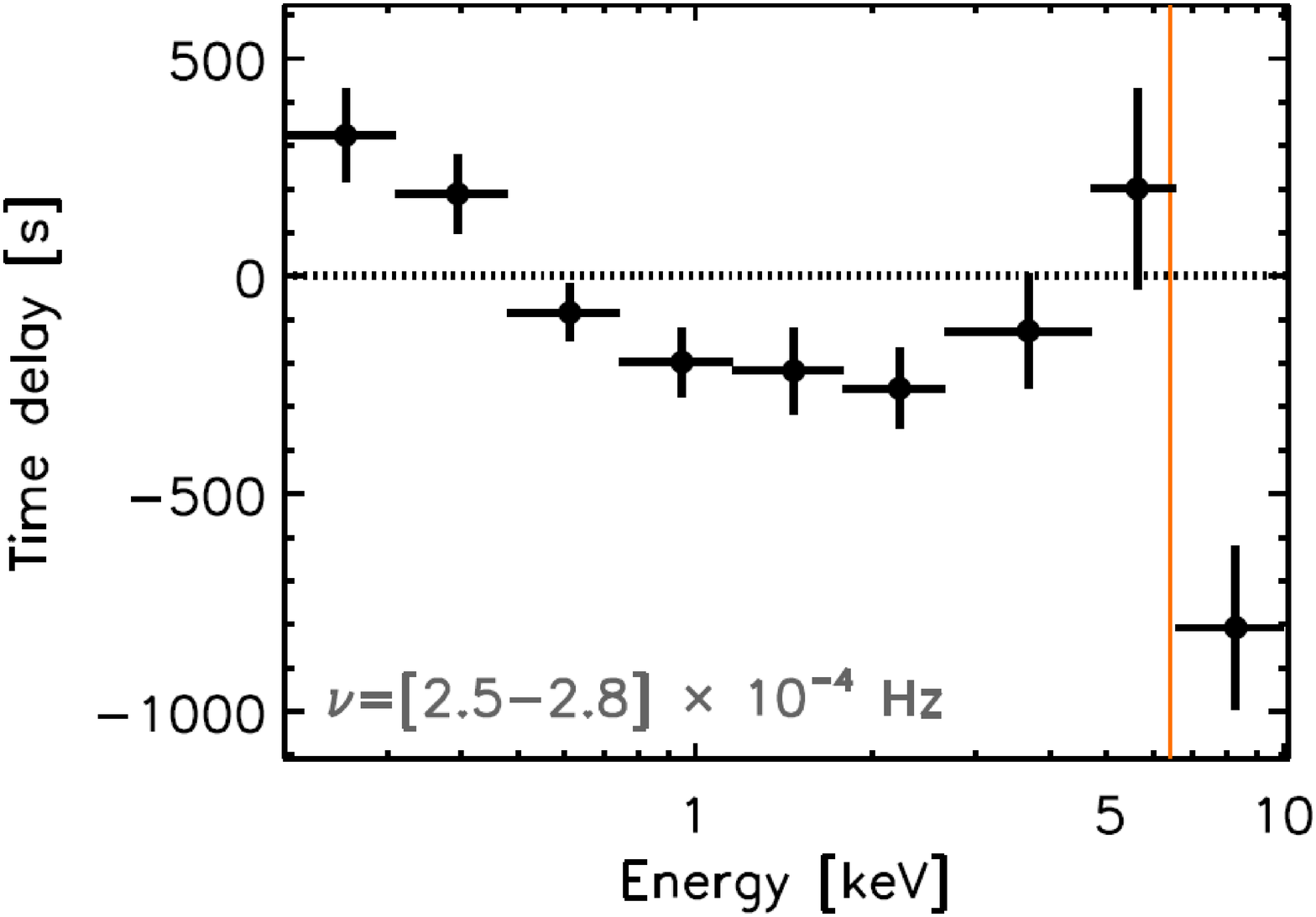}}
\caption{Time delays in \rej.  The top panel shows the time delay as a function of Fourier frequency between the hard ($1.0 - 4.0$\,keV) and soft ($0.3 - 0.8$\,keV) bands. Positive values indicate a hard band lag.  A hard lag is observed at low frequencies, which switches to a soft (negative) lag at the QPO freqeuncy ($\sim 2.6 \times 10^{-4}$\,Hz).  The bottom panel shows the lag-energy spectrum on the QPO frequency. A soft lag is observed as well as a lag in the iron K$\alpha$ band.  Figures to appear in Markeviciute et al {\it in prep}.}
\label{fig:rejlags}
\end{figure}
\begin{figure}
\centering
\subfigure{\includegraphics[width=0.44\textwidth,angle=0]{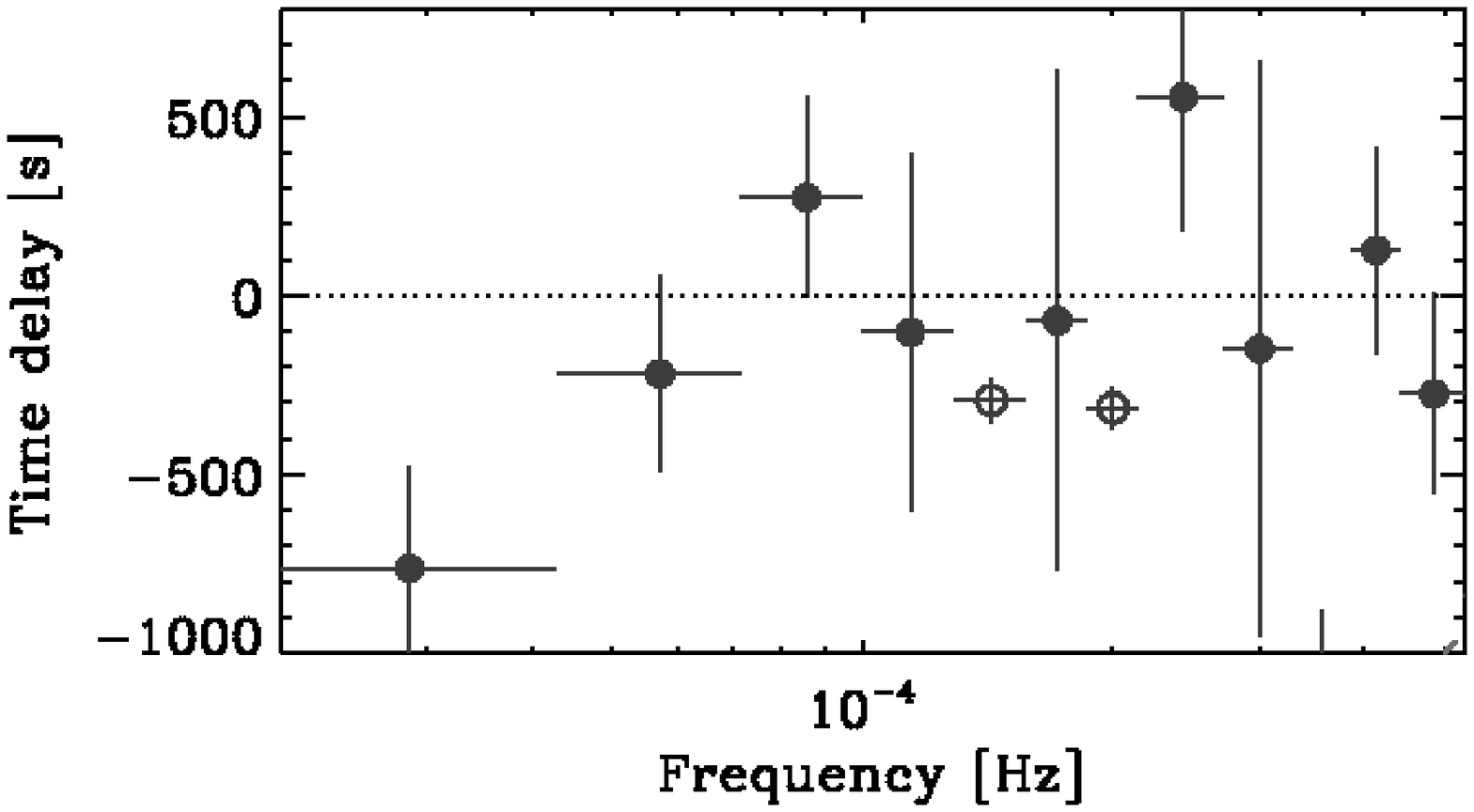}}
\\
\subfigure{\includegraphics[width=0.44\textwidth,angle=0]{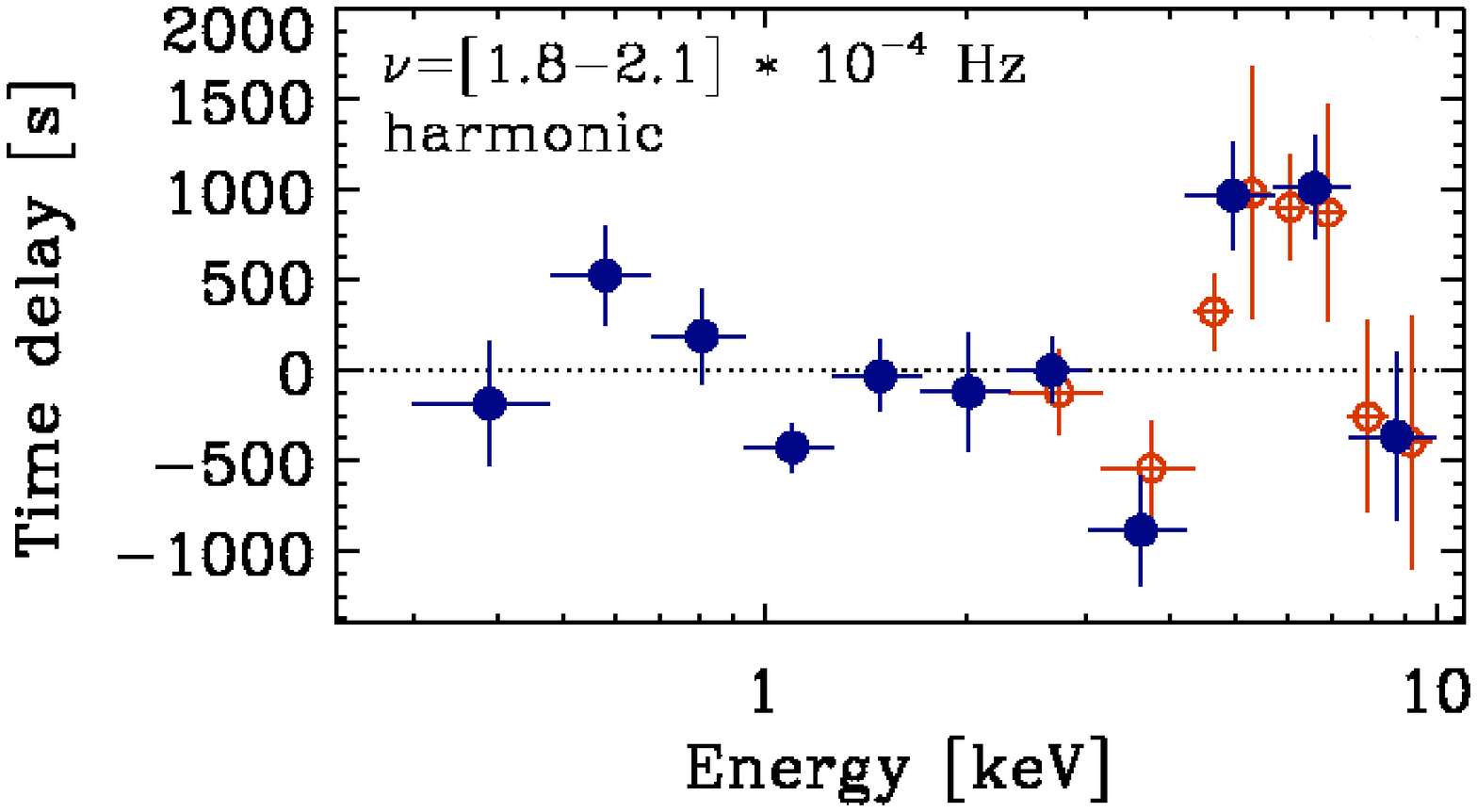}}
\caption{Time delays in \ms.  The top panel shows the time delay as a function of Fourier frequency between the hard ($1.2 - 5.0$\,keV) and soft ($0.3 - 0.7$\,keV) bands.  A soft (negative) lag is observed at at the QPO freqeuncy ($\sim 1.5 \times 10^{-4}$\,Hz) and harmonic frequency ($\sim 2 \times 10^{-4}$\,Hz).  The bottom panel shows the lag-energy spectrum on the harmonic frequency, where a lag in the iron K$\alpha$ band is observed.  Figures reproduced from A15.}
\label{fig:mslags}
\end{figure}

Following Vaughan \& Nowak (1997) we calculate the cross-spectrum in $M$ non-overlapping time series segments, then average over the $M$ estimates at each Fourier frequency.  Cross-spectra were averaged over adjacent frequency bins geometrically by a factor 1.15 in frequency.  From the complex valued cross-spectrum we obtain a phase lag estimate at each frequency, $\phi(f)$, which is gives the corresponding time lag $\tau(f) = \phi(f) / (2 \pi f)$.  Errors are estimated following e.g. Bendat \& Piersol (1986); Vaughan \& Nowak (1997).

For \rej~we make use of only the 5 observations where the QPO is detected (A14; Markeviciute et al, {\it in prep}).  Fig.~\ref{fig:rejlags} (top panel) shows the frequency dependent time lags between the $0.3-0.8$\,keV and $1.0-4.0$\,keV bands.  A positive lags indicates the hard band is lagging.  At low frequencies a hard lag is observed, whereas at the QPO frequency, a negative (soft) lag is observed (see also Zogbhi et al 2011).

A related technique is the \emph{lag-energy} spectrum: time delays at a given frequency as a function of energy.  The lag-energy spectrum is calculated by estimating the time delay between a comparison energy band vs a broad (in energy) reference band (e.g. Zoghbi 2011; Alston et al 2014a; Uttley et al 2014).  When the comparison energy band falls within the reference band that light curve is subtracted from the reference band, in order to avoid correlated errors.  Fig.~\ref{fig:rejlags} (bottom panel) shows the lag-energy spectrum for \rej~on the QPO timescale, where a positive lag indicates a given energy band lags the reference band (Markeviciute et al {\it in prep}).  The whole $0.3 - 10.0$\,keV band is used as the reference band.  A $\sim 500$\,s soft lag can be seen as well as a lag $\sim 1000$\,s at energy bands around the 6.4\,keV iron K$\alpha$ feature.

Fig.~\ref{fig:mslags} (top panel) shows the frequency dependent time lags between the $0.3-0.7$\,keV and $1.2-5.0$\,keV bands in \ms~(see Alston et al 2015 for more details).  A significant soft lag is seen at the QPO frequency, and at a frequency in an approximately 3:2 ratio with the QPO (open symbols).  No significant hard lag is observed in \ms~on the timescales we can probe, but a $\sim 3 \sigma$ soft lag is observed at the lowest frequency.  

The lag energy-spectrum for the QPO frequency in \ms~shows a soft lag, but no lag at iron K$\alpha$ (see Fig. 6 in A15).  The lag-energy spectrum for the harmonic frequency is shown in Fig.~\ref{fig:mslags} (bottom panel).  The $0.3 - 5.0$\,keV band is used as the reference.  A lag of $\sim 1000$\,s is observed between the continuum dominated bands and the $\sim 5-7$\,keV band, which contains the iron K$\alpha$ band.

\section{Conclusions}

\subsection{Summary}

We have presented recent work on QPOs in two AGN: \rej~and \ms, including investigation of their time delays. The results are summarised as follows:

\begin{itemize}
\item Using the Bayesian posterior predictive method of Vaughan (2012) we assess various continuum models to the observed data.  We then test for the presence of any interesting additional features where a low $p$-value is taken as evidence for an additional feature.  Significant narrow outliers are considered as evidence for the presence of a QPO.\\
\vspace{-0.3cm}
\item The QPO at $\sim 2.6 \times 10^{-4}$\,Hz ($\sim 1$\,hrs) in \rej~has been detected in four new \xmmn observations, spanning $\sim 5$ years.  The feature is only detected in the hard band ($1.0 - 4.0$\,keV) in observations that are lower in flux and are spectrally harder.  This increases the observation time where a QPO is detected in this source to $\sim 250$\,ks. There is tentative evidence for a harmonic component in one observation (Obs 7).\\
\vspace{-0.3cm}
\item A new QPO at $\sim 1.5 \times 10^{-4}$\,Hz ($\sim 2$\,hrs) is detected in the NLS1 galaxy, \ms.  Again, the feature is only detected in the continuum dominated band, $1.2 - 5.0$\,keV.\\
\vspace{-0.3cm}
\item The fractional variability of the features are 5 and 6 percent and quality factor, $Q \gsim 10$ and 8 in \rej~and \ms, respectively.\\
\vspace{-0.3cm}
\item The two sources share many of the same variability properties, including rms-spectra and principle components analysis (see A14 and A15 for more details).\\
\vspace{-0.3cm}
\item Investigation of the time delays between energy bands as a function of Fourier frequency reveals {\emph soft} lags at the QPO frequency (and harmonic frequency in \ms).  Iron K$\alpha$ like lags are seen at the QPO frequency in \rej, but only the harmonic frequency in \ms.\\
\vspace{-0.3cm}
\end{itemize}

\subsection{Discussion}

The central black hole mass of \rej~is uncertain, with best estimates are putting it at $\sim 1-4 \times 10^{6} \Msun$ (e.g. Bian \& Huang 2010; Jin et al 2011).  Based on this mass estimate and the inferred $L/\Ledd \gsim 1$, M10 and M11 proposed the QPO in \rej~was analogous with the 67 mHz QPO in GRS 1915+105.  However, Mendez et al 2013 observed a soft-lag in the 35 mHz HFQPO in GRS 1915+105 (Belloni \& Altamirano 2013).  A soft lag is observed at the QPO frequency (see also Zoghbi et al 2011), leading Mendez et al 2013 to suggest the QPO in \rej~is an analogue of the 35 mHz QPO in GRS 1915+105.  The potential harmonic feature we see in Obs7 of \rej~(Fig.~\ref{fig:rejpsd}) could then be the analogue of the 67 mHz QPO in GRS 1915+105.

The mass of the BH in \ms~is highly uncertain, with no reverberation mapping mass measurement fo rthis source.  In A15 we argue that the QPO is more consistent with the HFQPO type seen in XRBs.  HFQPOs in XRBs have a typical fractional rms $\sim 5$ per cent (Remillard \& McClintock 2006), consistent with the 6 per cent value observed here.  A $Q \gsim 2$ is observed in HFQPOs in XRBs (e.g. Casella et al 2004), again consistent with our value of $Q \sim 8$.  HFQPOs in XRBs often display harmonic components, with an integer ratio of 3:2 (e.g. Remillard et al 2002, 2003; Remillard \& McClintock 2006).  Strong evidence for the presence of a 3:2 harmonic component is observed in \ms, suggesting this is a HFQPO. 

The origin of HFQPOs is still highly uncertain, but it is clear that it physical mechanism occurring in the direct vicinity of the BH.  If indeed we are seeing iron K$\alpha$ reverberation responding to the QPO process, this will allow us to understand both the QPO mechanism in better detail, and provide important constraints for any model for the HFQPO mechanism.  Whether these are the same kind of lagging mechanism that is seen in other Seyferts (e.g. Uttley et al 2014) remains an open question.
\\

\acknowledgements
WNA, ACF and EK acknowledge support from the European Union Seventh Framework Programme (FP7/2013--2017) under grant agreement n.312789, StrongGravity.  This paper is based on observations obtained with \xmm, an ESA science mission with instruments and contributions directly funded by ESA Member States and the USA (NASA).




\begin{thebibliography}{}


\bibitem[Abramowicz 
\& Klu{\'z}niak(2001)]{2001A&A...374L..19A} Abramowicz, M.~A., \& Klu{\'z}niak, W.\ 2001, \aap, 374, L19

\bibitem[Alston et al.(2013)]{2013MNRAS.435.1511A} Alston, W.~N., Vaughan, 
S., \& Uttley, P.\ 2013, \mnras, 435, 1511
\bibitem[Alston et al.(2014a)]{2014MNRAS.439.1548A} Alston, W.~N., Done, C., 
\& Vaughan, S.\ 2014a, \mnras, 439, 1548 
\bibitem[Alston et al.(2014b)]{2014MNRAS.445L..16A} Alston, W.~N., 
Markevi{\v c}i{\= u}t{\.e}, J., Kara, E., Fabian, A.~C., 
\& Middleton, M.\ 2014b, \mnras, 445, L16 (A14)
\bibitem[Alston et al.(2015)]{2015MNRAS.449..467A} Alston, W.~N., Parker, 
M.~L., Markevi{\v c}i{\= u}t{\.e}, J., et al.\ 2015, \mnras, 449, 467 (A15)
\bibitem[Alston et al., {\it in prep}]{alstoninprep} Alston, W.~N., et al. {\it in prep}

\bibitem[Markevi{\v c}i{\= u}t{\.e}, J., et al. {\it in prep}]{markevinprep} Markevi{\v c}i{\= u}t{\.e}, J., Alston, W.~N., et al. {\it in prep}


\bibitem[Altamirano \& Belloni(2012)]{2012ApJ...747L...4A} Altamirano, D., \& Belloni, T.\ 2012, \apjl, 747, L4

\bibitem[Ar{\'e}valo 
\& Uttley(2006)]{2006MNRAS.367..801A} Ar{\'e}valo, P., \& Uttley, P.\ 2006, \mnras, 367, 801
\bibitem[Belloni et al.(2012)]{2012MNRAS.426.1701B} Belloni, T.~M., Sanna, 
A., \& M{\'e}ndez, M.\ 2012, \mnras, 426, 1701
\bibitem[Belloni \& Altamirano(2013)]{2013MNRAS.432...19B} Belloni, T.~M., \& Altamirano, D.\ 2013a, \mnras, 432, 19
\bibitem[Belloni 
\& Altamirano(2013)]{2013MNRAS.432...10B} Belloni, T.~M., \& Altamirano, D.\ 2013b, \mnras, 432, 10
\bibitem[Bian \& Huang(2010)]{2010MNRAS.401..507B} Bian, W.-H., \& Huang, K.\ 2010, \mnras, 401, 507
\bibitem[Cackett et al.(2013)]{2013ApJ...764L...9C} Cackett, E.~M., Fabian, 
A.~C., Zogbhi, A., et al.\ 2013, \apjl, 764, L9
\bibitem[Casella et 
al.(2004)]{2004A&A...426..587C} Casella, P., Belloni, T., Homan, J., \& Stella, L.\ 2004, \aap, 426, 587
\bibitem[Czerny et al.(2010)]{2010A&A...524A..26C} Czerny, B., Lachowicz, P., Dov{\v c}iak, M., et al.\ 2010, \aap, 524, A26

\bibitem[Czerny et al.(2012)]{2012JPhCS.372a2055C} Czerny, B., Lachowicz, P., Dov{\v c}iak, M., et al.\ 2012, Journal of Physics Conference Series, 372, 012055 
\bibitem[Das 
\& Czerny(2011)]{2011MNRAS.414..627D} Das, T.~K., \& Czerny, B.\ 2011, \mnras, 414, 627
\bibitem[De Marco et al.(2013)]{2013MNRAS.431.2441D} De Marco, B., Ponti, 
G., Cappi, M., et al.\ 2013, \mnras, 431, 2441
\bibitem[Emmanoulopoulos et al.(2011)]{2011MNRAS.416L..94E} 
Emmanoulopoulos, D., McHardy, I.~M., 
\& Papadakis, I.~E.\ 2011, \mnras, 416, L94
\bibitem[Fabian et al.(2009)]{2009Natur.459..540F} Fabian, A.~C., Zoghbi, 
A., Ross, R.~R., et al.\ 2009, \nat, 459, 540
\bibitem[Gierli{\'n}ski et al.(2008)]{2008Natur.455..369G} Gierli{\'n}ski, M., Middleton, M., Ward, M., \& Done, C.\ 2008, \nat, 455, 369

\bibitem[Gonz{\'a}lez-Mart{\'{\i}}n \& Vaughan(2012)]{2012A&A...544A..80G} Gonz{\'a}lez-Mart{\'{\i}}n, O., \& Vaughan, S.\ 2012, \aap, 544, A80

\bibitem[Grupe et al.(2004)]{2004AJ....127..156G} Grupe, D., Wills, B.~J., 
Leighly, K.~M., \& Meusinger, H.\ 2004, \aj, 127, 156
\bibitem[Grupe et al.(2010)]{2010ApJS..187...64G} Grupe, D., Komossa, S., 
Leighly, K.~M., \& Page, K.~L.\ 2010, \apjs, 187, 64


\bibitem[Hu et al.(2014)]{2014ApJ...788...31H} Hu, C.-P., Chou, Y., Yang, T.-C., \& Su, Y.-H.\ 2014, \apj, 788, 31

\bibitem[Jin et al.(2012)]{2012MNRAS.420.1825J} Jin, C., Ward, M., Done, C., \& Gelbord, J.\ 2012, \mnras, 420, 1825

\bibitem[Kara et al.(2013)]{2013MNRAS.428.2795K} Kara, E., Fabian, A.~C., 
Cackett, E.~M., et al.\ 2013, \mnras, 428, 2795
\bibitem[Kara et al.(2013)]{2013MNRAS.434.1129K} Kara, E., Fabian, A.~C., 
Cackett, E.~M., et al.\ 2013, \mnras, 434, 1129
\bibitem[Kara et al.(2015)]{2015MNRAS.446..737K} Kara, E., Zoghbi, A., 
Marinucci, A., et al.\ 2015, \mnras, 446, 737


\bibitem[Markevi{\v c}i{\= u}t{\.e}, J., et al. {\it in prep}]{markeviciuteinprep} Markevi{\v c}i{\= u}t{\.e}, J., Alston, W.~N., et al. {\it in prep}
\bibitem[Markowitz et al.(2003)]{2003ApJ...593...96M} Markowitz, A., Edelson, R., Vaughan, S., et al.\ 2003, \apj, 593, 96

\bibitem[McHardy et al.(2004)]{2004MNRAS.348..783M} McHardy, I.~M., Papadakis, I.~E., Uttley, P., Page, M.~J., \& Mason, K.~O.\ 2004, \mnras, 348, 783
\bibitem[M{\'e}ndez et al.(2013)]{2013MNRAS.435.2132M} M{\'e}ndez, M., Altamirano, D., Belloni, T., \& Sanna, A.\ 2013, \mnras, 435, 2132
\bibitem[Middleton \& Done(2010)]{2010MNRAS.403....9M} Middleton, M., \& Done, C.\ 2010, \mnras, 403, 9 (M10)

\bibitem[Middleton et al.(2009)]{2009MNRAS.394..250M} Middleton, M., Done, C., Ward, M., Gierli{\'n}ski, M., \& Schurch, N.\ 2009, \mnras, 394, 250

\bibitem[Middleton et al.(2011)]{2011MNRAS.417..250M} Middleton, M., Uttley, P., \& Done, C.\ 2011, \mnras, 417, 250 (M11)

\bibitem[Parker et al.(2014)]{2014MNRAS.437..721P} Parker, M.~L., 
Marinucci, A., Brenneman, L., et al.\ 2014, \mnras, 437, 721
\bibitem[Parker et al.(2015)]{2015MNRAS.447...72P} Parker, M.~L., Fabian, 
A.~C., Matt, G., et al.\ 2015, \mnras, 447, 72 

\bibitem[Remillard \& McClintock(2006)]{2006ARA&A..44...49R} Remillard, R.~A., \& McClintock, J.~E.\ 2006, \araa, 44, 49

\bibitem[Rezzolla et al.(2003)]{2003MNRAS.344L..37R} Rezzolla, L., Yoshida, 
S., Maccarone, T.~J., \& Zanotti, O.\ 2003, \mnras, 344, L37
\bibitem[Shields et al.(2003)]{2003ApJ...583..124S} Shields, G.~A., 
Gebhardt, K., Salviander, S., et al.\ 2003, \apj, 583, 124
\bibitem[Stella et al.(1999)]{1999ApJ...524L..63S} Stella, L., Vietri, M., 
\& Morsink, S.~M.\ 1999, \apjl, 524, L63
\bibitem[Strohmayer(2001)]{2001ApJ...552L..49S} Strohmayer, T.~E.\ 2001, \apjl, 552, L49

\bibitem[Uttley et al.(2002)]{2002MNRAS.332..231U} Uttley, P., McHardy, 
I.~M., \& Papadakis, I.~E.\ 2002, \mnras, 332, 231
\bibitem[Uttley et 
al.(2014)]{2014A&ARv..22...72U} Uttley, P., Cackett, E.~M., Fabian, A.~C., Kara, E., \& Wilkins, D.~R.\ 2014, \aapr, 22, 72

\bibitem[Vaughan 
\& Fabian(2003)]{2003MNRAS.341..496V} Vaughan, S., \& Fabian, A.~C.\ 2003, \mnras, 341, 496
\bibitem[Vaughan et al.(2003)]{2003MNRAS.345.1271V} Vaughan, S., Edelson, 
R., Warwick, R.~S., \& Uttley, P.\ 2003, \mnras, 345, 1271
\bibitem[Vaughan et al.(2003)]{2003MNRAS.339.1237V} Vaughan, S., Fabian, 
A.~C., \& Nandra, K.\ 2003, \mnras, 339, 1237
\bibitem[Vaughan(2005)]{2005A&A...431..391V} Vaughan, S.\ 2005, \aap, 431, 391
\bibitem[Vaughan 
\& Uttley(2005)]{2005MNRAS.362..235V} Vaughan, S., \& Uttley, P.\ 2005, \mnras, 362, 235
\bibitem[Vaughan 
\& Uttley(2006)]{2006AdSpR..38.1405V} Vaughan, S., \& Uttley, P.\ 2006, Advances in Space Research, 38, 1405
\bibitem[Vaughan(2010)]{2010MNRAS.402..307V} Vaughan, S.\ 2010, \mnras, 
402, 307
\bibitem[Vaughan et al.(2011)]{2011MNRAS.413.2489V} Vaughan, S., Uttley, 
P., Pounds, K.~A., Nandra, K., 
\& Strohmayer, T.~E.\ 2011, \mnras, 413, 2489

\bibitem[Wagoner(1999)]{1999PhR...311..259W} Wagoner, R.~V.\ 1999, 
\physrep, 311, 259
\bibitem[Wilkinson 
\& Uttley(2009)]{2009MNRAS.397..666W} Wilkinson, T., \& Uttley, P.\ 2009, \mnras, 397, 666

\bibitem[Zoghbi 
\& Fabian(2011)]{2011MNRAS.418.2642Z} Zoghbi, A., \& Fabian, A.~C.\ 2011, \mnras, 418, 2642
\bibitem[Zoghbi et al.(2012)]{2012MNRAS.422..129Z} Zoghbi, A., Fabian, 
A.~C., Reynolds, C.~S., \& Cackett, E.~M.\ 2012, \mnras, 422, 129
\bibitem[Zoghbi et al.(2014)]{2014ApJ...789...56Z} Zoghbi, A., Cackett, 
E.~M., Reynolds, C., et al.\ 2014, \apj, 789, 56

\end{thebibliography}
\end{document}